\documentclass[
    reprint, 
    aps,
    preprintnumbers,
    twocolumn,
    prb,
    superscriptaddress
]{revtex4-2}

\usepackage[utf8]{inputenc}
\usepackage{booktabs}
\usepackage[multiple]{footmisc}
\usepackage{lipsum}
\usepackage{rotating}
\usepackage{perpage}
\usepackage{chronology}
\usepackage{lipsum}

\usepackage{amssymb}
\usepackage{amsbsy}
\usepackage{amsmath}
\usepackage{abraces}
\usepackage{siunitx}
\usepackage{physics}
\usepackage[version=4]{mhchem}
\usepackage{bm}
\usepackage{bbm}
\usepackage{xfrac}
\usepackage{mathtools}

\usepackage{tikz}
\usepackage[T1]{fontenc}
\usepackage{etoolbox}
\usepackage{graphics}
\usepackage{multirow}
\usepackage{url}
\usepackage{hyperref}
\usepackage{placeins}
\usepackage[normalem]{ulem}
\usepackage[autostyle]{csquotes}

\usepackage{color}

\usepackage{cleveref}
\crefname{section}{Sec.}{Secs.}
\crefname{figure}{Fig.}{Figs.}
\crefname{table}{Tab.}{Tabs.}
\crefname{equation}{Eq.}{Eqs.}
\crefname{appendix}{App.}{Apps.}
\def\refcite#1{Ref.~\cite{#1}} 

\def\dul#1{\underline{\underline{#1}}}
\def\enforce#1{$\text{#1}$}

\newcommand{\mb}{\mathbf}

\def\usmash#1{^{\smash[t]{#1}}}
\def\dsmash#1{_{\smash[t]{#1}}}

\def\motionalav#1{\overline{#1}^{\text{ma}}}
\def\meanfieldav#1{\overline{#1}^{\text{mf}}}
\def\powderav#1{\overline{#1}^{\text{pa}}}

\newcommand{\Sx}{\mb{S}^{x}}
\newcommand{\Sy}{\mb{S}^{y}}
\newcommand{\Sz}{\mb{S}^{z}}
\newcommand{\Sxi}{\Sx_{i}}
\newcommand{\Syi}{\Sy_{i}}
\newcommand{\Szi}{\Sz_{i}}
\newcommand{\Sxj}{\Sx_{j}}
\newcommand{\Syj}{\Sy_{j}}
\newcommand{\Szj}{\Sz_{j}}

\newcommand{\Iy}{\mb{I}^{y}}
\newcommand{\Iz}{\mb{I}^{z}}

\newcommand{\vS}{\vec{\mb{S}}}

\newcommand{\Fl}{\mathrm{F}}
\newcommand{\Hy}{\mathrm{H}}
\newcommand{\HA}{\mathrm{HA}}
\newcommand{\HB}{\mathrm{HB}}
\newcommand{\Ca}{\mathrm{C}}
\newcommand{\CA}{\mathrm{CA}}
\newcommand{\CB}{\mathrm{CB}}

\newcommand{\Q}{\mathrm{Q}}

\newcommand{\JL}{J_{\mathrm{L}}}
\newcommand{\JQ}{J_{\Q}}
\newcommand{\JT}{J_{\mathrm{T}}}

\newcommand{\textJQ}{J\dsmash{\mathrm{Q}}}

\newcommand{\afcc}{a_{\text{fcc}}}
\newcommand{\asc}{a_{\text{sc}}}
\newcommand{\zeff}{z_{\text{eff}}}
\def\zeffof#1{z_{\text{eff}}^{\scriptscriptstyle#1}}
\newcommand{\nB}{\vec{n}_{B}}

\begin{document} 

\title{Microscopic understanding of NMR signals by dynamic mean-field theory for spins}

\def\shapeplotwidth{0.85\columnwidth}

\author{Timo Gr\"a{\ss}er}
\email{timo.graesser@tu-dortmund.de}
\affiliation{Condensed Matter Theory, TU Dortmund University,
Otto-Hahn Stra\ss{}e 4, 44221 Dortmund, Germany}

\author{Thomas Hahn}
\email{thomas.hahn@student.manchester.ac.uk}
\affiliation{Condensed Matter Theory, TU Dortmund University,
Otto-Hahn Stra\ss{}e 4, 44221 Dortmund, Germany}
\affiliation{School of Physics and Astronomy, The University of Manchester, Manchester M13 9PL, United Kingdom}

\author{G\"otz S.~Uhrig}
\email{goetz.uhrig@tu-dortmund.de}
\affiliation{Condensed Matter Theory, TU Dortmund University,
Otto-Hahn Stra\ss{}e 4, 44221 Dortmund, Germany}

\date{\today}

\begin{abstract}
A recently developed dynamic mean-field theory for disordered spins (spinDMFT) is shown to capture the spin dynamics of nuclear spins very well. The key quantities are the spin autocorrelations. In order to compute the free induction decay (FID), pair correlations are needed in addition. They can be computed on spin clusters of moderate size which are coupled to the dynamic mean fields determined in a first step by spinDMFT. We dub this versatile approach non-local spinDMFT (nl-spinDMFT). It is a particular asset of nl-spinDMFT that one knows from where the contributions to the FID stem. We illustrate the strengths of nl-spinDMFT in comparison to experimental data for CaF$_2$. Furthermore, spinDMFT provides the dynamic mean fields explaining the FID of the nuclear spins of $^{13}$C in adamantane up to some static noise. The spin Hahn echo in adamantane is free from effects of static noise and agrees excellently with the spinDMFT results without further fitting.
\end{abstract}

\maketitle

\section{Introduction}
\label{s:intro}

Nuclear magnetic resonance (NMR) has been successfully employed for decades to obtain microscopic information about matter \cite{levit05}. 
Experimental measurements of free-induction decays (FIDs), spin echoes and more allow conclusions to be drawn about chemical compounds and molecular distances in a sample. At the same time, a fully theoretical description of the measured quantities is rarely possible, since it requires the simulation of large numbers of nuclear spins. Conventional methods such as exact diagonalization and Chebyshev polynomial expansion are restricted to a couple of tens of spins \cite{talez84,weiss06a}. To capture realistic system sizes, fundamental approximations have to be considered. For the computation of FIDs, many different approximation schemes have been considered in the past \cite{vleck48,lowe57,parke74,fine97,stark18,stark20}(see also the references cited in \refcite{stark18}). Although some of them are quite accurate for the considered example, they are mostly difficult to generalize and, thus, rarely used at present. Recently proposed strategies include replacing all or many of the spins by classical ones \cite{elsay15,stark18,stark20}, which is particularly successful in systems where each spin has a large number of neighbors. The predictive power of such approaches can be questioned if quantum effects are more important, for instance, due to strong local quantum bonds or quadrupolar interactions \cite{glazo18}. Turning to non-equilibrium NMR measurements makes the situation even more complicated. A non-Markovian simulation of spin echoes, for example, is rarely even attempted \cite{fine05,witze05,witze06}.

The concept of a ``mean field'' has been well established in physics and chemistry for many decades. Static mean-field theories are not known for their accuracy, but they often provide us with a first qualitative picture of a complex many-body problem. Making the mean fields dynamic immensely increases the accuracy, see the fermionic DMFT developed in the nineties \cite{georg96} and its various extensions, for example cellular or extended DMFT \cite{kotli01,kotli02}. The recently developed dynamic mean-field theory for spins (spinDMFT) \cite{graes21} can be seen as an analogue for spin systems. Yet, spinDMFT is derived only for disordered spin systems. This makes it less general, but perfectly tailored to NMR. 

The main purpose of this article is to establish spinDMFT and its extensions in the NMR community as simulation tools. In doing so, we present an extension of spinDMFT which we call non-local spinDMFT (nl-spinDMFT). The idea is to include the mean-field moments computed by spinDMFT in the simulation of a local quantum cluster coupled to stochastic mean fields. This approach allows for an efficient computation of \emph{non-local} quantities and is related to the cluster-extension of spinDMFT (CspinDMFT) \cite{graes23}. 

We emphasize two advantages of these approaches. The first one is that the effective mean-field model maintains a microscopic picture of the considered system. Quantities such as spin-spin correlations are directly accessible which gives us a deeper insight into how the global NMR measurement is connected to microscopic spin physics. We will demonstrate this for the example of calcium fluoride, which is a common test substance for benchmarking approximation schemes due to a relatively noise-free FID in experiment \cite{engel74}. The second advantage of spinDMFT is its high versatility. Treating several spins on the quantum level as it is done in CspinDMFT can be very helpful in inhomogeneous systems \cite{graes23}. For spins with $S>\sfrac12$, a quadrupolar interaction can be directly included in the effective mean-field model. Moreover, external time-dependent fields and noises are easily included in spinDMFT \emph{and} its extensions without increasing the numerical effort much. Therefore, dynamical decoupling \cite{uhrig07,ryan10,alvar10c,ajoy11,alvar11b,souza11,souza12} can also be simulated within these approaches. We will demonstrate this in the second part of the article for the example of polycrystalline adamantane, which is another common test substance in NMR \cite{levit05,alvar10c,ajoy11,alvar11b,souza11,souza12}. 

\cref{sec:model} is devoted to a short overview of the considered substances and a brief derivation of spinDMFT. Subsequently, we introduce nl-spinDMFT and use it to estimate the FID of calcium fluoride in \cref{sec:CaFFID}. In \cref{sec:AdaProton}, we consider the microscopic proton-spin dynamics in adamantane using spinDMFT. The spin dynamics are used in \cref{sec:AdaCarbon} to compute the carbon FID and spin echo in adamantane.

\section{Models and \enforce{spin}DMFT}
\label{sec:model}

\subsection{Calcium fluoride}

Calcium fluoride, $\ce{CaF}_{2}$, is a crystal with calcium ions on an fcc-lattice and fluoride ions ordered in an sc-lattice intertwined with the former. For NMR purposes, the calcium ions can be neglected, since the abundance of nuclei with nonzero spin is very small, $\approx \SI{0.13}{\percent}$. A well-measurable NMR signal results from fluoride $\ce{^{19}F}$ which has an abundance of $\SI{100}{\percent}$ and carries a nuclear spin of $S~=~\sfrac12$. 

The basis for an NMR experiment is a strong magnetic field $\vec{B}$, which polarizes the spin ensemble into the field direction $\nB \coloneqq \vec{B}/|\vec{B}|$. The secular Hamiltonian for the homonuclear dipole-dipole interaction is given by 
\begin{align}
    \mb{H}_{\text{hom}} &= \frac12 \sum_{i,j} d_{ij} \left( - \Sxi \Sxj - \Syi \Syj + 2\Szi \Szj \right).
    \label{eqn:homonuclear}%
\end{align}
Here and henceforth, we set $d_{ii}=0$. The factor $\sfrac12$ in front of the sum accounts for double counting of the couplings. If $i\neq j$, the secular couplings read
\begin{subequations}
\begin{align}
    d_{ij} &\coloneqq d_{\vec{r}_{ij}}(\nB) = \frac{1 - 3 \left(\vec{n}_{ij}\cdot\nB\right)^2}{2} \frac{\mu_0}{4\pi} \frac{\gamma_i \gamma_j \hbar^2}{|\vec{r}_{ij}|^3}, \\
    \vec{r}_{ij} &\coloneqq \vec{r}_{i} - \vec{r}_{j} \qquad \vec{n}_{ij} \coloneqq \vec{r}_{ij}/|\vec{r}_{ij}|, 
\end{align}
\label{eqn:secularcouplings}%
\end{subequations}
with $\vec{r}_{i}$ and $\vec{r}_{j}$ being the nuclear spin positions fixed by the lattice and $\gamma_i$ and $\gamma_j$ being the gyromagnetic ratios. The latter are equal in the homonuclear case. Note that we shifted the $\hbar$'s into the couplings, that is, the spin vectors do not have a unit anymore.

\subsection{Adamantane}

\begin{figure}
    \centering
    \includegraphics[width=\columnwidth]{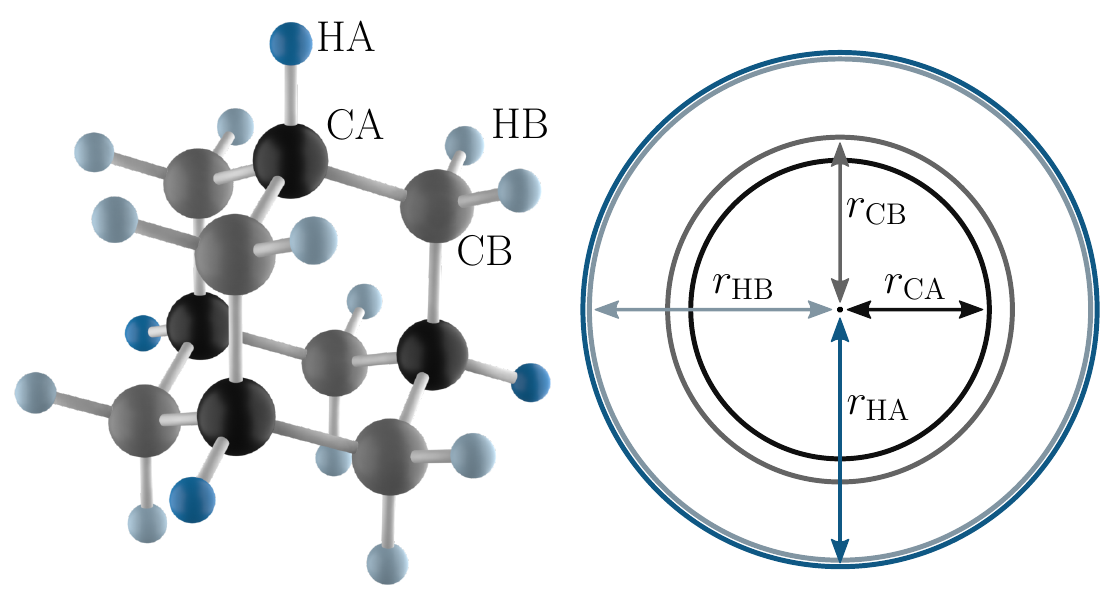}
    \caption{Sketch of the adamantane molecule. The hydrogen atoms are shown in dark blue ($\HA$) and light blue ($\HB$) and the carbon atoms in black ($\CA$) and gray ($\CB$). We use circles with the corresponding colors to represent the radii of the atoms with respect to the center of mass of the molecule, see \cref{eqn:radii}. The circles for the two types of hydrogen are nearly indistinguishable.}
    \label{fig:adamantane}
\end{figure}

The adamantane molecule consists of 10 carbon and 16 hydrogen atoms located on approximate spheres. Four of the carbon atoms ($\CA$) are each bound to one hydrogen atom ($\HA$) and six carbons ($\CB$)
are each bound to two hydrogen atoms ($\HB$). The molecule is displayed in the left panel of \cref{fig:adamantane}. We set all bond angles to the tetrahedral angle $\alpha~\approx~\SI{109.5}{\degree}$ and use the bond lengths \cite{hargi71}
\begin{align}
    r_{\Ca-\Ca} &= \SI{1.540(2)}{\angstrom}, & r_{\Ca-\Hy} &= \SI{1.112(4)}{\angstrom}.
\end{align}
The distances between the four types of atoms with respect to the center of mass of the molecule are then given by
\begin{subequations}
\begin{align}
    r_\CA &\approx \SI{1.54}{\angstrom}, & r_\CB &\approx \SI{1.78}{\angstrom}, \\
    r_\HA &\approx \SI{2.65}{\angstrom}, & r_\HB &\approx \SI{2.58}{\angstrom}.
\end{align}
\label{eqn:radii}%
\end{subequations}
Considering these values and \cref{fig:adamantane}, we note that the two classes of hydrogen and the two classes of carbon are each located nearly on the same shell. Hence, we consider only a single type of hydrogen and a single type of carbon at the averaged radii 
\begin{subequations}
\begin{align}
    r_{\Hy} &\coloneqq \frac{12 r_{\HA} + 4 r_{\HB}}{16} = \SI{2.60(3)}{\angstrom}, \\
    r_{\Ca} &\coloneqq \frac{6 r_{\CA} + 4 r_{\CB}}{10} = \SI{1.68(12)}{\angstrom}
\end{align}
\label{eqn:nucleusCMS}%
\end{subequations}
henceforth. As we will see later, this simplifies the application of spinDMFT, since only a single kind of mean field needs to be considered.

In line with the experiment in  \refcite{alvar10c}, we consider polycrystalline adamantane at room temperature. The adamantane molecules form an fcc-lattice with $\afcc~=~\SI{9.426(8)}{\angstrom}$ and four molecules per unit cell \cite{fort64}. While their centers of mass are fixed, the molecules are constantly changing their orientations. These molecular rotations are isotropic and happen on a much shorter time scale than the spin dynamics so that motional narrowing has to be considered for the couplings.

Aside from the proton spins, only the carbon isotopes $\ce{^{13}C}$ are of interest for NMR, since they carry a spin of $I~=~\sfrac12$ in contrast to the spinless nuclei of $\ce{^{12}C}$. Generally, we denote the $\ce{^{1}H}$ nuclear spins by $S$ and the $\ce{^{13}C}$ nuclear spins by $I$. The natural abundance of $\ce{^{13}C}$ is about $\SI{1.1}{\percent}$. Since the carbon spins are so rare, we neglect any 
carbon-hydrogen interaction for the hydrogen dynamics and any carbon-carbon interaction in general. The homonuclear secular Hamiltonian is given by
\begin{align}
    \mb{H}_{\text{hom}} &= \frac12 \sum_{i,j} \motionalav{d_{ij}} \left( - \Sxi \Sxj - \Syi \Syj + 2\Szi \Szj \right).
    \label{eqn:homonuclear2}
\end{align}
This is equivalent to \cref{eqn:homonuclear}, but with the couplings $J_{ij}$ replaced by their motional average $\motionalav{d_{ij}}$, see below. For the dynamics of $\ce{^{13}C}$, we also use the heteronuclear Hamiltonian
\begin{align}
    \mb{H}_{\text{het}} &= \sum_{j} \motionalav{d_{0j}} \, 2 \Iz \Szj,
    \label{eqn:carbondyn}
\end{align}
where, again, the averaged couplings have to be considered.
The indices $i \in \{1,2,3,\dots\}$ label the proton spins and the carbon index is 0.
As justified above, we consider the $\ce{^{13}C}$ spins to be isolated from one another.

The motional averaging causes any intramolecular couplings to average out completely, i.e., motional narrowing takes place. The averaged intermolecular couplings depend on (i) the distances of the considered nuclei to the center of mass and (ii) the lattice vector between the molecules to which the nuclei belong. For this reason, it is helpful to introduce double indices $(m X)\coloneqq i$, where $m \in \{1,2,3,\dots\}$ 
denotes the molecule and $X\in \Hy, \Ca$ is the nucleus type. The averaged couplings can be computed from
\begin{subequations}
\begin{align}
    \motionalav{d_{ij}}(\vec{n}_{B}) \coloneqq &\motionalav{d_{(m_1 X_1),(m_2 X_2)}}(\vec{n}_{B}) \\
    = \int \frac{\mathrm{d}\Omega_1 \mathrm{d}\Omega_2}{(4\pi)^2}& \, d_{\vec{a}_{m_1,m_2} + r_{X_1}\vec{n}(\Omega_{1}) - r_{X_2}\vec{n}(\Omega_{2})}(\vec{n}_{B}).
\label{eqn:motionalav}%
\end{align}
\end{subequations}
Here, $\vec{a}_{m_1,m_2}$ is the lattice vector from center to center of the molecules and 
\begin{align}
    \vec{n}(\Omega) &= \vec{n}(\theta,\phi) \coloneqq \begin{pmatrix}\sin\vartheta \cos\varphi \\ \sin\vartheta\sin\varphi \\ \cos\vartheta) \\ \end{pmatrix}. 
\end{align}
The radii $r_{X_1}, r_{X_2}$ are the distances from the nucleus to the center of its molecule, see \cref{eqn:nucleusCMS}.

An exact many-body treatment of adamantane is even further beyond reach than for calcium fluoride. Due to the motional averaging the intermolecular couplings are the dominant ones. Note that a single molecule already carries $16$ proton spins, so the number of couplings is immense, so that brute-force calculations are out of question. On the other hand, this situation is ideal for a mean-field approach. Due to the large number of interaction partners, the mean-field average is dominant over fluctuations. In the following, we present a brief summary of spinDMFT and show the generic results.

\subsection{Simulation of homonuclear dynamics by \enforce{spin}DMFT}
\label{subsec:spinDMFT}

Applying spinDMFT to a system containing only a single kind of spins is straight-forward \cite{graes21}. As an example, we consider the homonuclear Hamiltonian in \cref{eqn:homonuclear}. Defining the operators of the local environment, 
\begin{align}
    \vec{\mb{V}}_{i} &\coloneqq \sum_{j} d_{ij} \dul{D} \, \vS_{j}, & \dul{D} &= 
    \begin{pmatrix}
        -1 & 0 & 0 \\
        0 & -1 & 0 \\        
        0 & 0 & 2 \\
    \end{pmatrix},
    \label{eqn:locenv}
\end{align}
it can be rewritten according to 
\begin{align}
    \mb{H}_{\text{hom}} &= \frac12 \sum_{i} \vS_{i} \cdot \vec{\mb{V}}_{i}.
\end{align}
Note that the easy-axis anisotropy is entirely hidden in the operators $\vec{\mb{V}}_{i}$. The first step of spinDMFT is to replace these operators by classical time-dependent mean fields $\vec{V}_{i}(t)$. Each mean field summarizes the interaction of a large number of spins. Therefore, according to the central limit theorem, the mean fields follow a Gaussian distribution. Replacing the local-environment operators, we arrive at an effective single-site model 
\begin{align}
    \mb{H}_{i}^{\text{mf}}(t) &= \vS_{i} \cdot \vec{V}_{i}(t).
    \label{eqn:meanfieldham}
\end{align}
The quantum degrees of freedom of each spin are decoupled from those of the rest of the ensemble; hence, the spin index can be omitted henceforth. The mean fields can vary from site to site, even on average. However, for the fluoride ions in $\ce{CaF}_{2}$ as well as for the protons in adamantane, translational invariance is a well justified assumption. Therefore, the distribution of each mean field is independent of the site $i$ so that this index can be omitted as well. We consider a single site, i.e., a single spin and a single mean field henceforth.

Since the mean field is Gaussian, only the first and the second moment are required to fully determine it. In the high-temperature regime of completely disordered spins, expectation values are evaluated with respect to the disordered state $\rho_0~\propto~\mathbbm{1}$. The first moment is thus zero, 
\begin{align}
    \meanfieldav{V^{\alpha}(t)} &= \langle \mb{V}^{\alpha}(t) \rangle = 0,
\end{align}
where $\alpha \in \{x,y,z\}$ and the line with upper index ``mf'' denotes the mean-field averaging. The second moments can be related to quantum correlations of the original field according to
\begin{subequations}
\begin{align}
    \meanfieldav{V^{\alpha}(t)V^{\alpha}(0)} &= \langle \mb{V}^{\alpha}(t)\mb{V}^{\alpha}(0) \rangle \\
    &= \left(D^{\alpha\alpha}\right)^2 \JQ^2 \langle \mb{S}^{\alpha}(t)\mb{S}^{\alpha}(0) \rangle,
    \label{eqn:selfcons1} \\
    \meanfieldav{V^{\alpha}(t)V^{\beta}(0)} &= 0 \qquad \alpha\neq\beta.
    \label{eqn:selfcons2}
\end{align}
\label{eqn:selfcons}
\end{subequations}
where $D^{\alpha\alpha}$ is a diagonal matrix element of $\dul{D}$, see \cref{eqn:locenv}, and we defined the quadratic coupling constant
\begin{align}
    \JQ^2 \coloneqq \sum_{j} d_{ij}^2.
    \label{eqn:quadcoup}
\end{align}
Note that although the sum runs only over $j$, the total expression does not depend on $i$ due to translational invariance. We stress that correlations between mean fields at different sites are not relevant in spinDMFT due to the large coordination number \cite{graes21}. For the derivation of \cref{eqn:selfcons1} we used that pair correlations are subdominant and can be neglected relative to the autocorrelations, that is,
\begin{align}
    \sum_{\substack{j,k \\ j \neq k}} d_{ij} d_{ik} \langle \mb{S}_{j}^{\alpha}(t)\mb{S}_{k}^{\alpha}(0) \rangle
    &\ll \sum_{j} d_{ij}^2 \langle \mb{S}_{j}^{\alpha}(t)\mb{S}_{j}^{\alpha}(0) \rangle.
    \label{eqn:suppressedpairs}
\end{align}
This is also justified by the large coordination number $z$. Note that $z$ is typically defined by the number of nearest neighbors in the spin lattice. However, since the dipolar interaction is long-range, the effective number of interaction partners of each spin can be a lot higher than $z$. Therefore, we consider the 
effective coordination number\footnote{In \refcite{graes21}, we used the symbol $z_2$ for $\zeff$, because we discussed also $z_1 = \sfrac{\JL^2}{\JQ^2}$, where $\JL$ is the linear coupling sum.}
\begin{align}
    \zeff &\coloneqq \frac{\JQ^4}{\JT^4}, & \JT^4 &\coloneqq \sum_j d_{ij}^4,
\end{align}
to be more relevant for the justification of the mean-field approach. For the sake of brevity, we sometimes use $z$ instead of $\zeff$ in our argumentation.

\cref{eqn:meanfieldham,eqn:selfcons1,eqn:selfcons2} form a closed self-consistency problem which can be solved numerically. A comprehensive justification of spinDMFT and details on the numerical implementation can be found in the original article \cite{graes21}. In essence, one has to discretize the time equidistantly and simulate a single spin in a Gaussian noise, which can be done by Monte-Carlo simulation. The resulting spin autocorrelations are transformed to mean-field moments using the self-consistency conditions in \cref{eqn:selfcons1,eqn:selfcons2}, which are then fed into the next simulation of a single spin in a Gaussian noise. This process is repeated until the spin autocorrelations have converged which takes only about 5 iterations. We emphasize that the numerical errors from the Monte-Carlo simulation and the time discretization should always be tracked. The statistical error scales like $\sfrac1{\sqrt{M}}$, where $M$ is the sample size, and the discretization error typically like $\Delta t^2$, where $\Delta t$ is the time-step width.

\begin{figure}
    \centering
    \includegraphics{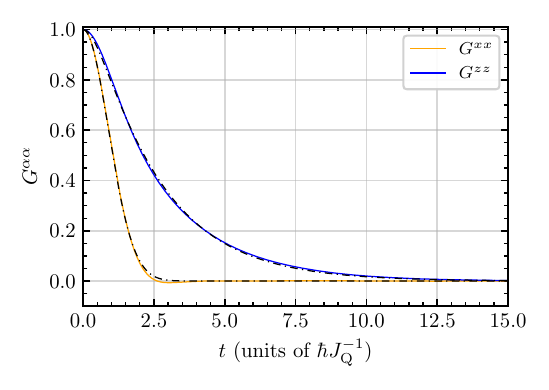}
    \caption{Generic spin autocorrelations computed by spinDMFT for dipolar spins in the rotating frame according to \cref{eqn:homonuclear}. The quadratic coupling constant is defined in \cref{eqn:quadcoup} and depends on the geometry of the spin system. The dashed-dotted black lines correspond to the fit functions in \cref{eqn:spinDMFTfits}. The sample size for the simulation is $M~=~\num{1e6}$ and the step width is $\Delta t~=~\num{1e-2}\,\hbar\textJQ^{-1}$ yielding an error of $\epsilon~<~2\times 10\usmash{-3}$.}
    \label{fig:generic}
\end{figure}

Henceforth, we will occasionally use the short-hands
\begin{subequations}
\begin{align}
    g_{ij}^{\alpha\alpha}(t) &\coloneqq \langle \mb{S}_{i}^{\alpha}(t)\mb{S}_{j}^{\alpha}(0) \rangle, \\
    G_{ij}^{\alpha\alpha}(t) &\coloneqq 4 g_{ij}^{\alpha\alpha}(t).
\end{align}
\end{subequations}
The site indices are omitted if $i=j$. One should note that the only remaining parameter in the self-consistency problem is the quadratic coupling constant $\JQ$. This constant sets the time scale. We present the generic autocorrelation results of spinDMFT for Hamiltonians of type \eqref{eqn:homonuclear} in \cref{fig:generic}. The difference in the time scale between the transverse ($G^{xx}$) and longitudinal ($G^{zz}$) autocorrelation is expected due to the anisotropy factor of 2 in the couplings. We find, that the decays are very well captured by the fits 
\begin{subequations}
\begin{align}
    F^{xx}(t) &= \text{exp}\left(-\frac12 \left(\frac{t}{\sigma} \right)^2\right)\\
    F^{zz}(t) &= \text{exp}\left(-\rho \left( \sqrt{t^2 + \kappa^2} - |\kappa| \right)\right)
\end{align}
\label{eqn:spinDMFTfits}
\end{subequations}
with
\begin{subequations}
\begin{align}
    \sigma &= \num{0.8891(4)}\,\hbar\JQ^{-1}, \\
    \rho &= \num{0.4326(6)}\,\hbar^{-1}\JQ, \\
    \kappa &= \num{0.647(4)}\,\hbar\JQ^{-1}.
\end{align}
\end{subequations}
We emphasize the qualitative difference in the temporal dependence of $F^{xx}$ and $F^{zz}$ for $t~\to~\infty$, Gaussian versus exponential. The spin autocorrelations yield a good estimate for the time scale of the spin dynamics. The key quantity is the quadratic coupling constant. 

\section{Simulating FID\enforce{s} with \enforce{nl-spin}DMFT}
\label{sec:CaFFID}

We stress that computing the FID with \enforce{spin}DMFT represents a challenging task, since the approach accesses the spin dynamics from a microscopic point of view, while the FID is a macroscopic quantity. Our strategy for computing the FID is to relate it to microscopic spin-spin correlations. This is explained in the first part of this section. Then, we present an extension of spinDMFT, called nl-spinDMFT, and use it to evaluate the FID of calcium fluoride.

\subsection{FID and microscopic correlations}

The FID corresponds to a measurement of the transverse magnetization over time considering a slightly polarized initial state with the statistical operator \cite{cho05}
\begin{align}
    \rho_{\rightarrow} &\coloneqq \frac1{Z} \mathrm{e}^{h \sum_{j} \mb{S}_{j}^{x}} \approx \frac1{Z} \left( \mathbbm{1} + h \sum_{j} \mb{S}_{j}^{x} \right)
\end{align}
where $h\coloneqq\beta\hbar\gamma B$, $\beta$ is the inverse temperature, $Z$ is the partition sum and $\gamma$ is the gyromagnetic ratio. The transverse magnetization can be expressed by infinite temperature spin correlations according to
\begin{align}
    \langle \mb{M}^{x}(t) \rangle_{\rho_{\rightarrow}} &= A_0 \sum_{i,j} \langle \mb{S}^{x}_{i}(t) \mb{S}^{x}_{j} \rangle, 
\end{align}
where $A_0$ is a time-independent prefactor, which is not relevant to us. The double sum can be divided into a sum over autocorrelations and a sum over pair correlations according to 
\begin{align}
    \underbrace{\sum_{i}}_{\propto N} \underbrace{\langle \mb{S}^{x}_{i}(t) \mb{S}^{x}_{i} \rangle}_{\propto 1} + \underbrace{\sum_{i}}_{\propto N}\underbrace{\sum_{j\neq i}}_{\propto z} \underbrace{\langle \mb{S}^{x}_{i}(t) \mb{S}^{x}_{j} \rangle}_{\leq\propto \sfrac{1}{z}}.
    \label{eqn:paircorrelations}
\end{align}
We recall that a key aspect of spinDMFT is that pair correlations are suppressed by at least a factor $\sfrac{1}{z}$, where $z$ is the coordination number \cite{zobov88,graes21}. Often, this allows us to neglect them for the dynamics, e.g., in \cref{eqn:suppressedpairs}. We stress that for this argument to work the presence of the couplings as factors in \cref{eqn:suppressedpairs} is crucial. They assist in suppressing any contribution from pair correlations. In \cref{eqn:paircorrelations}, however, the double sum for the pair correlations contains no couplings. If we consider only nearest-neighbor pair correlations, the double sum contains $\propto z N$ terms, where $N$ is the particle number. As a consequence, the whole expression scales like $N$ which is the same as the scaling of the autocorrelation sum. It is the additional factor $z$ which has the pair correlations play a role in the FID although they are unimportant for the microscopic spin dynamics. This is the key observation in this section.

Intuitively, one would expect the relevance of pair correlations to decrease as the direct couplings of the two involved spins decreases. We stress, however, that indirect couplings, for example via a third spin, can also matter. It is not \emph{a priori} clear, if or how much pair correlations become irrelevant if we increase the distance of the two involved spins. However, we will see that the more indirect a coupling is, the more time a correlation needs to build up between the corresponding spins. To capture the FID at short and moderate times, it is a promising strategy to start with the pair correlations of close spins. In \cref{tab:shelllist}, we list and number pair correlations, sorted first by the distance between the pair of spins, and second, if the distances are equal, by the coupling between the spins. 

\begin{table}[]
    \centering
    \begin{tabular}{|c||ccc|S[table-format=1.4]|S[table-format=1.4]|c|}
    \hline \\[-2.5ex] 
    $c$ & \multicolumn{3}{c|}{$\vec{r}_{c}^{\top}/\asc$} & $|\vec{r}_{c}|/\asc$ & {$|d_{c}|/d_{0}$} & $m_c$ \\ [0.5ex] 
    \hline
    0 &      & & &        &      &      1 \\      
    1 &      (0 & 0 & -1) &       1.0 &      2.0 &      2 \\      
    2 &      (-1 & 0 & 0) &       1.0 &      1.0 &      4 \\      
    3 &      (-1 & -1 & 0) &      1.4142 &   0.3535 &   4 \\      
    4 &      (-1 & 0 & -1) &      1.4142 &   0.1768 &   8 \\      
    5 &      (-1 & -1 & -1) &     1.732 &    0.0 &      8 \\      
    6 &      (0 & 0 & -2) &       2.0 &      0.25 &     2 \\      
    7 &      (-2 & 0 & 0) &       2.0 &      0.125 &    4 \\      
    8 &      (-1 & 0 & -2) &      2.2361 &   0.1252 &   8 \\      
    9 &      (-2 & -1 & 0) &      2.2361 &   0.0894 &   8 \\      
    10 &     (-2 & 0 & -1) &      2.2361 &   0.0358 &   8 \\      
    11 &     (-1 & -1 & -2) &     2.4495 &   0.068 &    8 \\      
    12 &     (-2 & -1 & -1) &     2.4495 &   0.034 &    16 \\     
    13 &     (-2 & -2 & 0) &      2.8284 &   0.0442 &   4 \\      
    14 &     (-2 & 0 & -2) &      2.8284 &   0.0221 &   8 \\      
    15 &     (0 & 0 & -3) &       3.0 &      0.0741 &   2 \\      
    16 &     (-3 & 0 & 0) &       3.0 &      0.037 &    4 \\      
    17 &     (-2 & -2 & -1) &     3.0 &      0.0247 &   8 \\      
    18 &     (-2 & -1 & -2) &     3.0 &      0.0124 &   16 \\     
    19 &     (-1 & 0 & -3) &      3.1623 &   0.0538 &   8 \\
    \hline
    \end{tabular}
    \caption{List of correlations considered for the FID of calcium fluoride in the [100]-direction ($z$). The first entry ($c=0$) is the autocorrelation and the other entries correspond to pair correlations. $\vec{r}_{c}$ is an exemplary distance vector between the two spins involved in the pair correlation and $\sfrac{|d_{c}|}{d_0}$ is the absolute of their direct couplings with $d_0 \coloneqq \mu_0 \gamma_{\Fl}^2 \hbar^2 / (4\pi\asc^3)$. By $m_c$ we denote the multiplicity of the correlation, i.e., the number how often this correlation occurs per spin.}
    \label{tab:shelllist}
\end{table}

Note that it is not possible to compute pair correlations in spinDMFT, since they are excluded from the approach by construction. The extension, CspinDMFT, exactly simulates a small number of spins, the so-called cluster, and regards the remainder of the lattice by mean fields \cite{graes23}. CspinDMFT can access these pair correlations, but each pair correlation would require an adapted cluster which would make the calculation numerically very expensive. In addition, the self-consistency in each cluster would yield autocorrelations which differ from cluster to cluster. This is not systematic. Thus we adopt the following two-step strategy instead.

\subsection{Overview of nl-spinDMFT}

\begin{figure}
    \centering
    \includegraphics[width=\columnwidth]{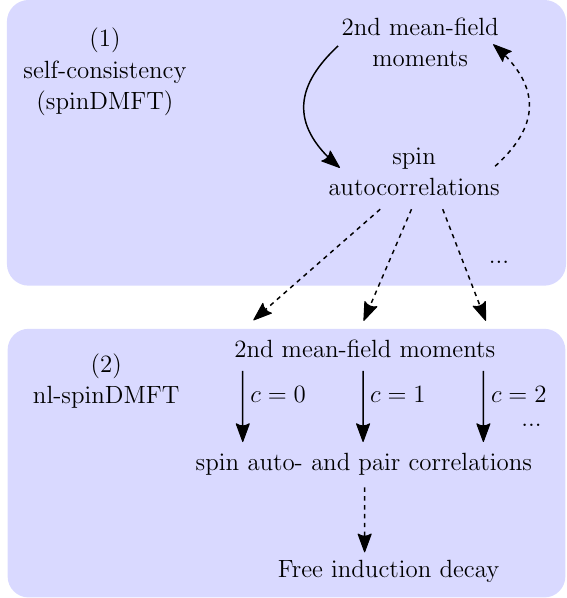}
    \caption{Strategy to compute spin correlations and finally the FID using dynamic mean-field theory. The dashed arrows correspond to simple, numerically cheap superpositions. The solid arrows correspond to numerically more demanding Monte-Carlo simulations which are required to compute expectation values.}
    \label{fig:FIDstrategy}
\end{figure}

In step (1), we estimate the autocorrelations by a single self-consistent computation with spinDMFT. This is well justified for large $\zeff$. In step (2), we perform simulations of a quantum cluster in a mean-field background to determine each pair correlation listed in \cref{tab:shelllist}. Thereby, we use the previously obtained autocorrelations as building blocks to determine the required mean-field moments. The simulations of step (2) are similar to CspinDMFT, but we do not require the self-consistency anymore because the mean-field moments are already well known from the upstream self-consistency in step (1). At the same time, we are flexible to choose the quantum cluster individually for each simulation in step (2). Hence, the choice of the cluster can be optimized with respect to the pair correlation of interest. Finally, the FID is obtained by superposing all correlations obtained in the second step. Since the pair correlations and the FID are non-local quantities, we refer to the procedure in second step as non-local spinDMFT (nl-spinDMFT). The whole strategy is summarized in \cref{fig:FIDstrategy}. We provide more details in the following.

The self-consistent calculation required for the first step is discussed in \cref{subsec:spinDMFT} (spinDMFT). If the coordination number is high, spinDMFT is sufficient, but for example in dimerized systems, one should simulate at least a single dimer of spins using CspinDMFT. For calcium fluoride with magnetic field in the [100]-direction, we have $\zeffof{[100]}~=~4.9$, which is sufficiently large for using only spinDMFT.

In the second step, we perform several simulations of a quantum cluster in a mean-field background. The cluster is henceforth denoted by $\Gamma_l$, where $l$ is the number of spins. The spins in the cluster interact quantum-mechanically with one another and we regard the remainder of the lattice by coupling each of the spins to a dynamic Gaussian mean field $\vec{V}_{i}$. This is summarized by the Hamiltonian 
\begin{align}
\begin{split}
    \mb{H}_{\Gamma_l}^{\text{mf}}(t) &= \frac12
		\sum_{i,j \in \Gamma_l} d_{ij} \left( - \Sxi \Sxj - \Syi \Syj + 2\Szi \Szj \right) \\
    &\qquad\qquad\qquad+\sum_{i \in \Gamma_l} \vec{\mb{S}}_{i} \cdot \vec{V}_{i}(t).
\end{split}
\label{eqn:multisitemodel}
\end{align}
As mentioned before, the second moments of the mean fields are determined from the autocorrelations obtained in the first step. Generally, the second moments read
\begin{align}
    \meanfieldav{V^{\alpha}_{i}(t)V^{\alpha}_{j}(0)} \!&= \left(D^{\alpha\alpha}\right)^2 \!\! \sum_{k,m\notin\Gamma_l} d_{ik} d_{jm} \langle \mb{S}_{k}^{\alpha}(t)\mb{S}_{m}^{\alpha}(0) \rangle. 
\label{eqn:mfmomentsCaFraw}
\end{align}
Using translational invariance and the sorting scheme from \cref{tab:shelllist}, this can be rewritten as 
\begin{subequations}
\begin{align}
    \meanfieldav{V^{\alpha}_{i}(t)V^{\alpha}_{j}(0)} &= \left(D^{\alpha\alpha}\right)^2 \sum_{c=0}^{\infty} J_{(c),ij}^2 G_{c}^{\alpha\alpha}(t), \\
    J_{(c),ij}^2 &\coloneqq \sum_{\substack{k,m\notin\Gamma_l \\ \vec{r}_k - \vec{r}_m \stackrel{!}{\in} \{\vec{r}_c\}}} d_{ik} d_{jm},
\end{align}
\label{eqn:mfmoments_reorder}%
\end{subequations}
where $G_{c}^{xx}$ is the pair correlation of type $c$ and $G_{0}^{xx}~=~G^{xx}$ is the autocorrelation. By $\{\vec{r}_c\}$ we denote the set of distance vectors $\vec{r}_c$ that correspond to the correlation with index $c$. The number of these distance vectors per site corresponds to the multiplicity $m_c$ which is also provided in \cref{tab:shelllist}. For example, the set for $c=1$ contains $\vec{r}_c = \begin{smallmatrix}(0 & 0 & -1)\end{smallmatrix}^{\top}$ and $\vec{r}_c = \begin{smallmatrix}(0 & 0 & 1)\end{smallmatrix}^{\top}$, which entails $m_c~=~2$. Of course, we can only insert those correlations that have been computed in the first step, i.e., the autocorrelations, so that the summation in \cref{eqn:mfmoments_reorder} reduces to
\begin{align}
    \meanfieldav{V^{\alpha}_{i}(t)V^{\alpha}_{j}(0)} &\approx \left(D^{\alpha\alpha}\right)^2 J_{(0),ij}^2 \langle \mb{S}^{\alpha}(t)\mb{S}^{\alpha}(0) \rangle.
\label{eqn:mfmoments_autotrunc}
\end{align}
Note that this approximation is also justified by the coordination number. For example, for $i=j$, the approximation in \cref{eqn:mfmoments_autotrunc} follows directly from \cref{eqn:suppressedpairs}.

The expression in \cref{eqn:mfmoments_autotrunc} can be straightforwardly computed after obtaining the autocorrelations in step (1). 
The resulting second mean-field moments are then fed into the simulations of a spin cluster $\Gamma_l$ in a mean-field background using the Hamiltonian in \cref{eqn:multisitemodel}. Note that expectation values are again evaluated with respect to the disordered state $\rho_0~\propto~\mathbbm{1}$. For calcium fluoride, we consider cluster sizes of up to $l=9$ spins and access all pair correlations listed in \cref{tab:shelllist}. Next, we discuss how one can optimize the choice of clusters for the computation of the pair correlations.

\subsection{Choice of the clusters}
\label{subsec:choiceofclusters}

 Clearly, for a pair correlation the two directly involved spins must be included in $\Gamma_l$. For a reasonable choice of the other spins, one has to consider that two spins can become correlated indirectly, for example, via their coupling to a third spin. Such processes can be only accounted for, if the third spin is included in $\Gamma_l$, since backactions from the cluster onto the mean fields are excluded by construction. For an optimal choice of the cluster, the most important indirect couplings have to be identified.

A selection criterion for the spins to be put into the cluster can be developed, for example, on the basis of the fourth moment of the coupling constants
\begin{align}
    \left(J_{i-k-j}\right)^4 &\coloneqq J_{ik}^2 J_{kj}^2.
    \label{eqn:JC4}
\end{align}
Here, $i$ and $j$ denote the pair of spins of the pair correlation in question and $k$ is some arbitrary third spin. The coupling constant connects spin $i$ and $j$ via two links, which captures an indirect process correlating the two spins with each other. We stress, however, that indirect processes involving three or more links can also be important. Respecting several degrees of indirect correlations simultaneously in a criterion forms a challenging task. 

Formally, we define an \emph{order} as the number of links between site $i$ and $j$. We propose a strategy that adds the presumably most important spins order by order to the cluster and start with the lowest relevant order, $n=2$. The first order, $n=1$, is automatically accounted for, since we always include the spins $i$ and $j$ in the cluster. Dipolar couplings are long-range so that, in principle, there is an infinite number of processes for $n\geq 2$. To develop an appropriate criterion, we consider only the $L$ strongest absolute couplings of a spin in the lattice. For example, for calium fluoride with magnetic field in the [100]-direction a plausible choice would be $L=6$, which includes only the nearest-neighbor couplings. We suggest the following procedure for the $n$-th order:
\begin{enumerate}
    \item Determine all paths from site $i$ to site $j$ with $n$ links. Only those links are allowed, that correspond to one of the $L$ strongest 
    couplings. 
    \item Collect the sites of all paths in a set $K$. Add all spins in $K$ to the cluster, if this does not exceed the desired cluster size $l$, and proceed with the next order.
    If adding all spins would exceed $l$, consider the next step instead.
    \item Let $f$ be the number of spins that are yet missing in the cluster. Find all possibilities to distribute $f$ spins on the sites in $K$.
    For each possibility, compute 
    \begin{align}
        \left(J_{i-\Sigma_n-j}\right)^{2n} \! (\Gamma_l) &\coloneqq \hspace{-0.65cm}\sum_{k_1, k_2, \dots, k_{n-1} \in \Gamma_l} \hspace{-0.65cm} d_{ik_1}^2 d_{k_1k_2}^2 \dots d_{k_{n-1}j}^2.
    \end{align}
    Here, $k_i$ can be any spin in the cluster with the $f$ spins added (total size is $l$) and we consider only the most important couplings. Accept the cluster $\Gamma_l$ with the highest value for $(J\dsmash{i-\Sigma_n-j})\usmash{2n}$ and stop the procedure.
\end{enumerate}
We consider this strategy to be a good compromise between lower and higher orders. Aside from the cluster size $l$ which is set by the available numerical resources, the only free parameter is the number of considered couplings $L$. 

\begin{figure}
  \centering
  \includegraphics[width=\shapeplotwidth]{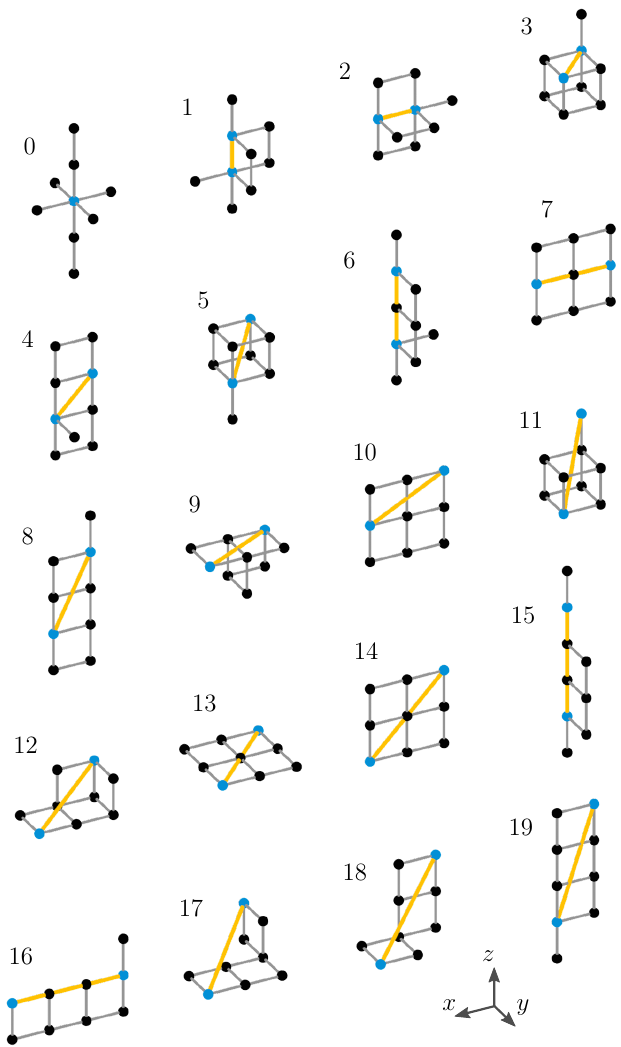}
  \caption{Considered clusters in the simple cubic lattice for $\ce{CaF_2}$ for magnetic field in the [100]-direction ($z$) according to the presented strategy in the text. The numbering is adapted from \cref{tab:shelllist}. The orange lines represent the pair correlations accessed by the corresponding cluster.}
  \label{fig:sh0-19}
\end{figure}

Carrying this procedure out for $L=6$, we arrive at the 20 clusters shown in \cref{fig:sh0-19}. Note that the procedure has also been used for determining the optimal cluster for the autocorrelation $c=0$ (spin $i=j$).

\subsection{Results for calcium fluoride}
\label{subsec:CaFresults}

\begin{figure*}
    \centering
    \includegraphics[width=\textwidth]{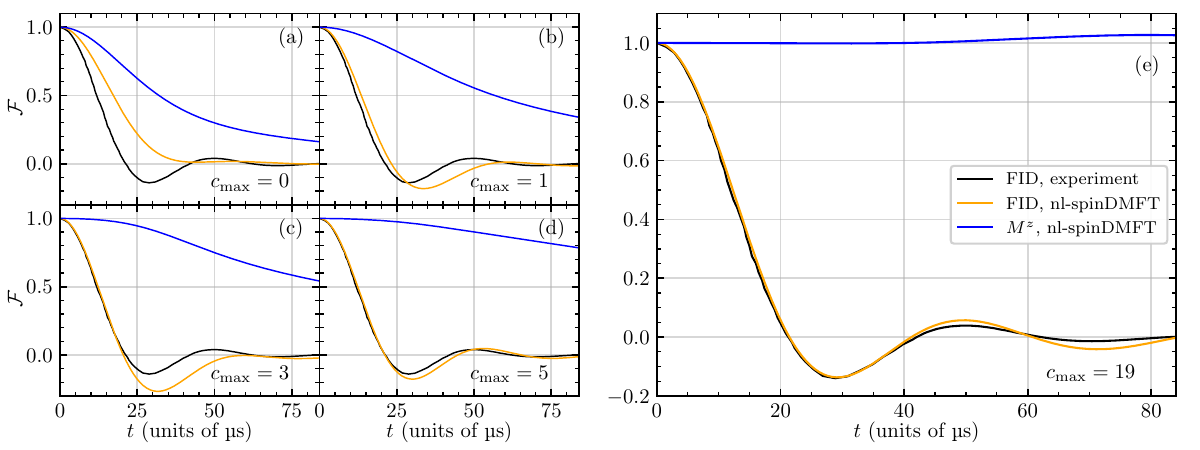}
    \caption{Comparison of the FID in the [100]-direction simulated with nl-spinDMFT with the experimental one measured in \refcite{engel74}. The clusters considered for the simulation are depicted in \cref{fig:sh0-19}; they comprise $l=9$ sites and were chosen considering the $L=6$ strongest couplings. The different panels show the improved agreement as $c_{\text{max}}$ is increased, see \cref{eqn:CaFFID} for clarity. In addition to the FID, we plot the longitudinal magnetization from nl-spinDMFT, see \cref{eqn:mz}. This quantity is rigorously conserved in the absence of any approximations. The individual correlations $G_{c}^{\alpha\alpha}$ were simulated considering a sample size of $M~=~\num{1e3}$ and a step width of $\Delta t~=~\SI{0.35}{\micro\second}$ yielding an error of $\epsilon~<~4\times 10\usmash{-3}$. The experimental data were measured at $\SI{4.2}{\kelvin}$. The lattice constant for the fluorine ions at this temperature is about $\asc~=~\SI{2.724}{\angstrom}$.}
    \label{fig:CaFFID100}
\end{figure*}

To obtain the actual contribution of a pair correlation to the FID $\mathcal{F}(t)$, one has to multiply it with its multiplicity $m_c$ in the lattice so that
\begin{align}
    \mathcal{F}(t) &= \sum_{c=0}^{\infty} m_{c} G_{c}^{xx}(t).
    \label{eqn:CaFFID}
\end{align}
In practice, the sum needs to be truncated. We denote the corresponding truncation value by $c_{\text{max}}$. The results of the simulated FID for ascending $c_{\text{max}}$ are plotted in \cref{fig:CaFFID100} in comparison with experimental data from \refcite{engel74}. In addition to the FID, we plot
\begin{align}
    M^{z}_{c_{\text{max}}}(t) &= \sum_{c=0}^{c_{\text{max}}} m_{c} G_{c}^{zz}(t),
    \label{eqn:mz}
\end{align}
which is exactly constant for $c_{\text{max}}~=~\infty$, since the longitudinal magnetization $\mb{M}^{z}$ is a conserved quantity. Hence, the deviation of $M^{z}(t)$ from $1$ is an excellent measure for the good accuracy of the approximation in \cref{fig:CaFFID100} (e).

As we increase the truncation $c_{\text{max}}$, the agreement between nl-spinDMFT and experiment improves as expected. Adding the contributions from 19 pair correlations leads to a remarkable agreement up to about $t=\SI{40}{\micro\second}$. Beyond this time, the amplitude of the oscillation is slightly overestimated by nl-spinDMFT. We can rule out the statistical error of the Monte-Carlo simulation as well as the error induced by time discretization as the origin of the deviation, since they are more than one order of magnitude smaller. A probable reason for the deviation is that the considered quantum clusters are not large enough to capture long-range pair correlations reliably. The latter result from indirect processes involving many spins and therefore become important at later times. 

\begin{figure}
    \centering
    \includegraphics[width=\columnwidth]{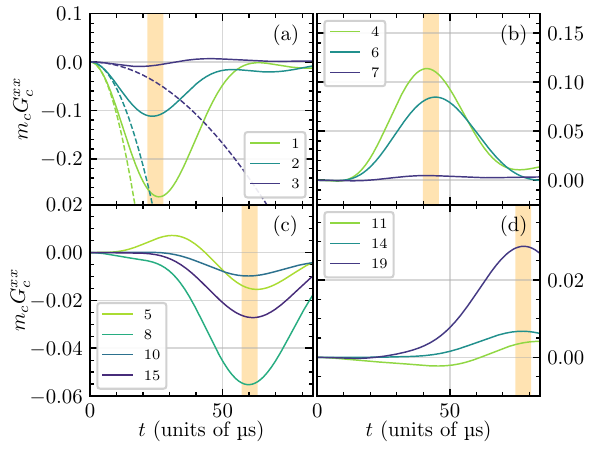}
    \caption{Contributions of the transverse pair correlations computed by nl-spinDMFT. The numbering corresponds to the one in \cref{tab:shelllist} and \cref{fig:sh0-19}. Some pair correlations are not shown as they are very close to zero (absolute value smaller than $0.002$ on the whole interval). The orange bars correspond to the approximate peak positions which are briefly discussed in the main text. The dashed lines result from \cref{eqn:dimercorr}.}
    \label{fig:CaFPairx}
\end{figure}

\begin{figure}
    \centering
    \includegraphics[width=\columnwidth]{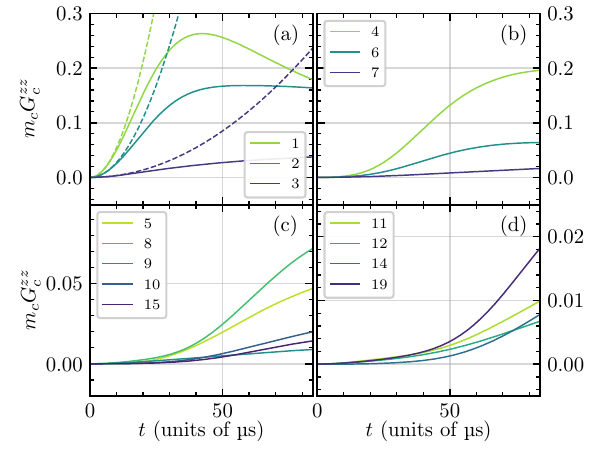}
    \caption{Same as \cref{fig:CaFPairx}, but for the longitudinal pair correlations. }
    \label{fig:CaFPairz}
\end{figure}

To illustrate this, we present the individual transverse and longitudinal pair correlations in \cref{fig:CaFPairx,fig:CaFPairz}. Some of them are very close to zero and, hence, are not shown. The transverse correlations are mostly characterized by a single peak, which can be positive or negative. Depending on the peak position and its sign we can group them phenomenologically into four classes on the considered time interval. The longitudinal pair correlations on the other hand are all positive. Most of them increase monotonically over the considered time interval. No.~1 and 2 rise the fastest and slowly decay after their maximum value at about $t=\SI{40}{\micro\second}$, see panel (a) in \cref{fig:CaFPairz}. Note that we applied the same classification scheme as for the transverse pair correlations. The speed of the increase of the longitudinal pair correlations decreases with $c$, i.e., with the pair distance.

Henceforth, we focus on the transverse pair correlations as they matter for the FID. The first peak is due to 1st-order processes, i.e., due to the direct coupling between the pair of spins of the corresponding pair correlation. We can verify this by analytically computing the exact pair correlations, while setting all couplings except the direct coupling $d$ to zero, yielding
\begin{subequations}
\begin{align}
    G^{xx}_{\text{dimer}}(t) &= - \sin\left(\frac{t d}{\hbar}\right) \sin\left(\frac{t d}{2\hbar}\right), \\
    G^{zz}_{\text{dimer}}(t) &= \sin^2\left(\frac{td}{2\hbar}\right),
\end{align}
\label{eqn:dimercorr}
\end{subequations}
which is depicted by the dashed lines in \cref{fig:CaFPairx,fig:CaFPairz} in panel (a), respectively. The agreement of these formulas with the pair correlations of nl-spinDMFT is excellent at short times which indicates that the direct coupling is the origin of the first peak in the transverse correlations. The other peaks are likely to result from indirect couplings. One can presume that the peak of the $n$th class results from processes of $n$th order according to the definition in \cref{subsec:choiceofclusters}, that is, from processes with $n$ links. This also explains the alternating signs of the peaks: If the direct coupling between two spins leads to anti-correlation ($-$) at short times, the indirect coupling via two links leads to correlation ($+$), the indirect coupling via three links again to anti-correlation and so on. Note that, generally, we expect all orders to be relevant for all pair correlations to some extent. This can be seen well in pair correlation No.~5 in panel (c), in which both the second and third peak are clearly visible. However, the obtained tendency is that the larger the distance between a pair of spins, the higher is the order of the processes inducing the relevant peak in the pair correlation. This entails that more-distant pair correlations tend to involve more links and, thus, tend to pop up at later times. In \cref{app:convergence}, we discuss the convergence of the correlations upon increasing the cluster size $l$. We find that short-range pair correlations converge much quicker than long-range pair correlations. This is plausible considering the fact that the order of the involved processes increases with the pair distance. 

Fig.~\ref{fig:CaFFID100} could give the impression that the mean-field approach corresponds to a short-time approximation, but this is not a valid view. In fact, nl-spinDMFT is a short-range approximation and range and time correspond to each other in the FID. We emphasize that short-ranged quantities such as the autocorrelations are well captured by the approach \emph{even at long times}. This is underlined by the consideration in \cref{app:convergence}.

There are several ways to improve the accuracy of the simulated FID at later times. An obvious step is to increase the cluster size further; one or two more spins are realistic. Besides this, one has to consider that the prime reason for inaccuracies in long-range pair correlations is the lack of back actions in the mean-field approach. In classical or hybrid quantum-classical approaches, such back actions are treated at least on a classical level \cite{elsay15,stark18,stark20}. 
The advantage of the expansion in \cref{eqn:CaFFID} is that the summed up quantities can all be computed separately. A promising strategy for very accurate FIDs could be to estimate any correlations below a properly chosen $c_{\mathrm{max}}$ by nl-spinDMFT and to compute correlations for $c>c_{\text{max}}$ by a fully classical or by a hybrid approach. We leave these ideas for future research.

In \cref{app:otherdir}, we also compare nl-spinDMFT with the experimental data for the FIDs in the [110]- and [111]-directions. The observations are qualitatively the same as above. The nl-spinDMFT is precise at short times and performs less well at later times, since long-range effects are not accounted for accurately. In fact, the deviations at later times are considerably higher than for the [100]-direction. We assign this to the fact, that the effective coordination number is much larger for the [110]- and [111]-directions, $\zeffof{[110]}~=~9.2$ and $\zeffof{[111]}~=~22.5$. This may sound counter intuitive, because the mean-field approximation is justified by a $\sfrac1{z}$-expansion. However, accessing pair correlations requires to simulate \emph{compact} clusters, in which the most important intermediate spins inducing indirect correlations are included. The larger the coordination number $z$, the larger is the number of important intermediate spins which complicates the choice of compact clusters enormously. If $z$ is lower, the mean-field approximation is, in principle, less well justified. However, the agreement of the results with the experimental FID in the [100]-direction together with the convergence plots in \cref{app:convergence} suggest that even for a smaller effective coordination number, $\zeffof{[100]}~=~4.9$, the local dynamics is captured accurately by the mean-field approach. Therefore, we believe that nl-spinDMFT can be a very powerful tool for the computation of FIDs in systems with moderate coordination numbers, for example, in dimerized systems such as crystalline silicon \cite{stark20} or in two-dimensional systems, see for example \refcite{graes23}. To access systems with very high coordination numbers, one may combine it with approaches that capture back actions more effectively.

\section{Microscopic proton-spin dynamics in adamantane}
\label{sec:AdaProton}

In the previous section, we demonstrated how spinDMFT and its extensions can be employed to simulate the FID. To this end, we computed individual microscopic pair correlations by simulating clusters of 9 spins in a mean-field background. An important ingredient was the choice of compact clusters on the lattice. Such a choice is not possible for adamantane, where the motional averaging leads to exceptionally high coordination numbers in the leading non-vanishing couplings to adjacent molecules and beyond. In this scenario, at least two or three molecules would need to be included for a reliable choice of clusters. This would entail simulating more than 30 proton spins which is numerically not feasible. We conclude on the one hand that the proton FID of adamantane cannot be simulated by the described procedure. On the other hand, the mean-field approximation itself is excellently justified for the nuclear spins in adamantane, even more so than for $\ce{CaF_2}$ because of the large number of interaction partners, i.e., large values of $z$. For this reason, we expect the autocorrelations of the proton spins to be captured very well by spinDMFT. Computing them requires powder averaging, which is discussed in the following.

To apply spinDMFT, we require the quadratic coupling constant for the proton spins. For a given direction $\vec{n}_B$ of the magnetic field, it can be computed by
\begin{align}
    J_{\Q,\Hy}^2(\vec{n}_{B}) &= N^{\text{mol}}_{\Hy} \sum_{m>1} \left(\motionalav{d_{(1 \Hy),(m \Hy)}}(\vec{n}_{B}) \right)^2,
    \label{eqn:JQH}
\end{align}
with the motion-averaged couplings provided in \cref{eqn:motionalav} and $N^{\text{mol}}_{\Hy}=16$ being the number of hydrogen nuclei per molecule.
To compute $J_{\Q,\Hy}$ numerically, we rewrite the molecule sum as a double sum and define a cutoff $s_{\text{max}}$,
\begin{align}
    J_{\text{Q},\Hy}^2(\vec{n}_{B}) &\approx N^{\text{mol}}_{\Hy} \sum_{s=1}^{s_{\text{max}}} \hspace{-0.05cm} \sum_{\substack{m \\ |\vec{a}_{1,m}| = r_s}} \hspace{-0.4cm}  \left(\motionalav{d_{(1 \Hy),(m \Hy)}}(\vec{n}_{B})\right)^2.
\end{align}
The first sum runs over all \enquote{shells} of molecules around the considered central molecule in the fcc lattice. The shells are defined by their distance to the central molecule. Here, $s_{\text{max}}$ is the index of the last shell that is included in the numerical calculation. The second sum runs over all molecules within the same shell, that means, with the same distance $r_s$ to the central molecule. The lattice distances $r_s$ and the number of molecules per shell in the fcc lattice are extracted from \refcite{burg11}. For an analytical estimate of the contributions from shells with $s>s_{\text{max}}$, we assume for simplicity each nucleus to be at the center of its molecule and introduce the continuum limit to render the molecule sum tractable. Both approximations become more accurate as $s_{\text{max}}$ is increased. We obtain
\begin{subequations}
\begin{align}
    J_{\text{Q},\Hy,\text{rest}}^2 &\approx N^{\text{mol}}_{\Hy} n_0 \int_{r_{s_{\text{max}}+1}}^{\infty} \hspace{-0.8cm} \mathrm{d}a \, a^2 \int \mathrm{d}\Omega \,d^2\hspace{-0.05cm}\left(a\vec{n}(\Omega),\vec{n}_{B}\right) \\
    &= \frac{16 \pi N^{\text{mol}}_{\Hy}}{15 a_{\text{fcc}}^3} \left( \frac{\mu_0 \gamma_{\Hy}^2 \hbar^2}{4\pi} \right)^2 (r_{s_{\text{max}}+1})^{-3},
    \label{eqn:JQrest}
\end{align}
\end{subequations}
where $\vec{n}(\Omega) = \vec{n}(\theta,\phi) \coloneqq \begin{smallmatrix}(\sin\vartheta \cos\varphi,& \sin\vartheta\sin\varphi, & \cos\vartheta)\end{smallmatrix}^{\top}$ and $n_0 \coloneqq 4/a_{\text{fcc}}^3$ is the density of molecules. Note that due to the continuum limit the lattice becomes completely isotropic so that the result is independent of $\vec{n}_B$. 

\begin{figure}
    \centering
    \includegraphics[width=\columnwidth]{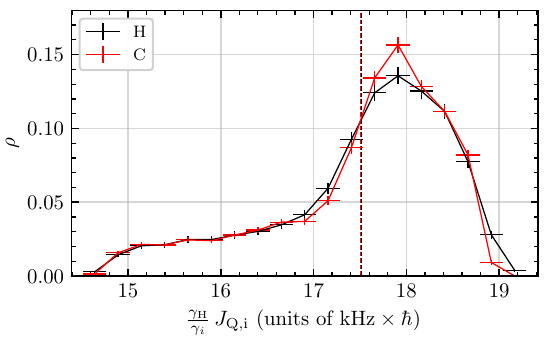}
    \caption{Distribution of the quadratic coupling constant $(\sfrac{\gamma_{\Hy}}{\gamma_{i}}) J_{\Q,i}$ for $i\in\{\Hy,\Ca\}$. The two dashed vertical lines correspond to the powder-average $\powderav{J_{\Q,i}}$ (they are barely distinguishable). $\rho$ is determined by organizing the resulting values for $J_{\Q,i}$ for different field directions $\vec{n}_B$ into bins. The horizontal error bars correspond to the bin widths. The vertical error bars are determined from the error bars of $J_{\Q,i}$ which result from the Monte Carlo sampling.}
    \label{fig:rhoJQ}
\end{figure}

\begin{figure}
    \centering
    \includegraphics[width=\columnwidth]{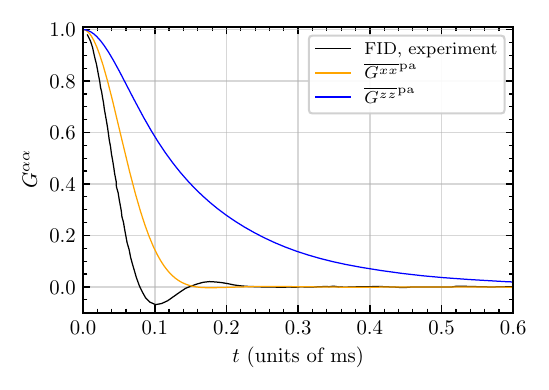}
    \caption{Powder-averaged proton autocorrelations for adamantane computed by spinDMFT in comparison with the proton FID measured in \refcite{alvar10c}. The agreement is only approximate because the pair correlations are missing.}
    \label{fig:protonAda}
\end{figure}

The coupling $\motionalav{d_{(1 \Hy),(m \Hy)}}$ has to be evaluated for each molecule $m$ in each shell $s$ for different orientations $\vec{n}_{B}$ of the magnetic field. We compute the four-dimensional angular integrals in \cref{eqn:motionalav} by Monte Carlo integration. In total, we consider the first eight shells around the central molecule, i.e. $s_{\text{max}} = 8$, and approximate contributions from higher shells using \cref{eqn:JQrest}. In spinDMFT, the orientation of the magnetic field itself is not important, but the resulting quadratic coupling $\JQ$ is the key parameter. Therefore, we are not interested in the distribution of $J_{\Q,\Hy}$ depending of $\vec{n}_{B}$, but rather in the distribution $\rho\left(J_{\Q,\Hy}\right)$, which is displayed in \cref{fig:rhoJQ}. The powder-averaged value from our simulation is $\powderav{J_{\Q,\Hy}}~=~\SI{17.519(3)}{\kilo\hertz}\times\hbar$ which agrees very well with the experimentally obtained value, $\powderav{J_{\Q,\Hy}}~\approx~\SI{17}{\kilo\hertz}\times\hbar$ \cite{tompa13} (\refcite{tompa13} provides the second moment $M_2$ which differs from the quadratic coupling constant according to $\JQ^2~=~M_2\hbar^2\gamma^2\sfrac{4}{9}$). However, as $\rho\left(J_{\Q,\Hy}\right)$ is not strongly peaked, but spreads over an interval of about $\SI{4}{\kilo\hertz}\times\hbar$, this average value is not completely representative. Applying the powder average directly to the quadratic coupling constant, that is, \emph{before} the spinDMFT simulation is not fully justified. Instead, we should produce spinDMFT results for different values of $J_{\Q,\Hy}$ and superpose the results according to the distribution $\rho\left(J_{\Q,\Hy}\right)$. Note that this does not require additional effort, since the simulation results from spinDMFT, see \cref{fig:generic}, can simply be rescaled in order to correspond to a different $J_{\Q,\Hy}$. The powder-averaged autocorrelations of spinDMFT obtained in this way are plotted in \cref{fig:protonAda} together with the experimentally measured FID. It turns out that the powder average does not affect the shapes of the autocorrelations much. 

In contrast to the FID, the transverse autocorrelation does not contain any oscillations. This is the same phenomenon as in calcium fluoride. The pure autocorrelation, see panel (a) of \cref{fig:CaFFID100}, does not oscillate, but after adding pair correlations the estimated FID contains minima and maxima, see panel (e). As explained above, we cannot access the pair correlations in adamantane reliably, since the required cluster sizes would be too large. Hence, there is no direct comparison between spinDMFT and experiment concerning the proton-spin dynamics. But the situation is much more favorable for the dynamics of the nuclear spins of $\ce{^{13}C}$, which will be considered in the next section.

\section{Simulating carbon NMR signals in adamantane}
\label{sec:AdaCarbon}

Since the $\ce{^{13}C}$-spins are rare, their dynamics are completely dominated by the interaction with the proton spins. The latter occur in much higher numbers and their gyromagnetic ratio is larger so that we can safely neglect any back action from the carbon spins onto the proton spins. In this case, a substitution of the proton field in \cref{eqn:carbondyn} by a Gaussian mean field is very justified. This leads to a simple spin-noise model according to the Hamiltonian
\begin{align}
    \mb{H}_{\Ca}(t) &= \Iz V^{z}_{\Ca}(t).
    \label{eqn:noiseham}
\end{align}
The mean field has zero average and its second moments are given by 
\begin{align}
    \meanfieldav{V^{z}_{\Ca}(t)V^{z}_{\Ca}(0)} &= 4 J_{\Q,\Ca}^2 \langle \Sz(t) \Sz(0) \rangle,
    \label{eqn:carbonmfmoment}
\end{align}
with the longitudinal spin autocorrelation provided in Fig.~\ref{fig:generic}.
Similar to \cref{eqn:JQH}, the quadratic coupling constant is given by 
\begin{align}
    J_{\Q,\Ca}^2(\vec{n}_{B}) &= N^{\text{mol}}_{\Hy} \sum_{m>1} \left(\motionalav{d_{(1 \Ca),(m \Hy)}}(\vec{n}_{B}) \right)^2.
\end{align}
Note that we do not consider the powder average in \cref{eqn:carbonmfmoment}. Within each crystal, the carbon spins couple to an individual spin environment with fixed $J_{\Q,\Ca}$ and fixed proton autocorrelation. Therefore, one has to compute any signals of the carbon spins individually for each crystal or, equivalently, for each $J_{\Q,\Ca}$ and the corresponding proton autocorrelation. Averaging these signals leads to the appropriate powder-average measured in experiment.

Using the same procedure as in the previous section, we can estimate $J_{\Q,\Ca}$ by Monte-Carlo integration. In our calculations it turns out that, although the carbon spins sit on a much smaller radius in the adamantane molecule, their quadratic coupling constant is always exactly the same as for the proton spins except for a factor $r_\gamma \coloneqq \sfrac{\gamma_{\Ca}}{\gamma_{\Hy}}\approx 0.251$, due to the different gyromagnetic ratios. This is true for any considered field direction $\vec{n}_B$. To underline this observation, we included the density $\rho(r_\gamma J_{\Q,\Ca})$ in \cref{fig:rhoJQ}, which is remarkably close to $\rho(J_{\Q,\Hy})$. Motivated by this, we also computed the distributions by a more crude model, where the nuclei are put to the centers of the molecules. Indeed, this changes the distributions only very slightly.

\begin{figure}
    \centering
    \includegraphics[width=\columnwidth]{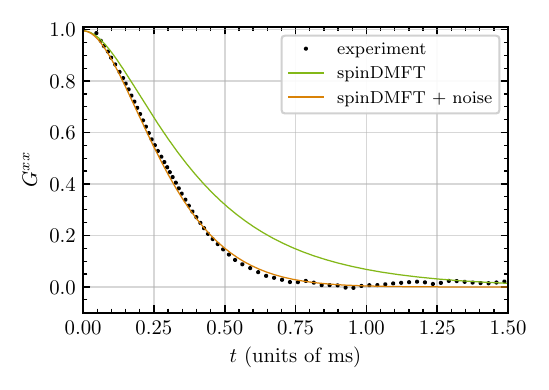}
    \caption{Powder-averaged transverse autocorrelation for $\ce{^{13}C}$ in adamantane from the spin-noise model in \cref{eqn:noiseham}. Results with and without an additional static noise source are shown. The carbon FID measured in \refcite{alvar10c} is plotted with black circles.}
    \label{fig:carbonFID}
\end{figure}

We compute the transverse autocorrelation generated by the Hamiltonian in \cref{eqn:noiseham} for different values of $J_{\Q,\Ca}$ and superpose the results according to $\rho(J_{\Q,\Ca})~\approx~\rho(r_\gamma^{-1} J_{\Q,\Hy})$ to obtain the powder average. The resulting signal is shown in Fig.~\ref{fig:carbonFID} together with the experimental result for the FID. Note that since the $\ce{^{13}C}$-atoms have a low abundance, two individual carbon spins see a totally different environment. Hence, pair correlations are not relevant for the corresponding FID and the transverse autocorrelation of carbon should, in principle, be equivalent to the FID. However, as it turns out, the result from the noise model deviates a bit from the experiment at moderate and large times. For a better understanding of this observation, we consider the spin echo signal in the following.

\begin{figure}
    \centering
    \includegraphics[width=\columnwidth]{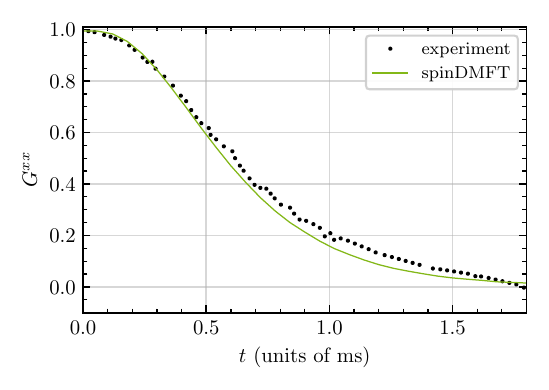}
    \caption{Powder-averaged spin echo for $\ce{^{13}C}$ in adamantane from the spin-noise model in \cref{eqn:noiseham}. The experimental spin echo measured in \refcite{alvar10c} is plotted with black circles.}
    \label{fig:spinecho}
\end{figure}

To simulate a spin Hahn echo \cite{hahn50} within the spin noise model, we include an instantaneous pulse according to
\begin{align}
    \mb{H}_{\text{pulse}} &= \pi \delta(t-\tau/2) \Iy
\end{align}
and measure the result of the transverse autocorrelation at $t=\tau$. The result of this is plotted versus $\tau$ in Fig.~\ref{fig:spinecho}. Here, the agreement between theory and experiment is very good. We conclude that the dynamic noise resulting from the proton spins as captured by spinDMFT is perfectly accurate. Since a spin echo eliminates static noise we attribute the discrepancy in the FID to some small additional static noise.

To underline this interpretation, we redo the computation of the FID and include a weak static Gaussian noise. This is done by adding a constant term $W^2$ to the second mean-field moment in \cref{eqn:carbonmfmoment}, where $W$ is the standard deviation of the noise. The best fit to the experimental data is obtained for $W = \SI{0.29(6)}{\kilo\hertz}\times\hbar$ and plotted in orange in Fig.~\ref{fig:carbonFID}. The agreement for the adapted model is remarkable and supports our assumption. We stress that the spin echo exactly reverts the effect of the noise in the considered model so that the result in Fig.~\ref{fig:spinecho} does not change at all. 

In addition, we analyzed how such an additional noise source would affect the proton-spin dynamics. To this end, we added the noise, amplified by the factor $\sfrac1{r_\gamma}$, in the spinDMFT simulation finding no visible difference in the resulting autocorrelations relative to the ones shown in Fig.~\ref{fig:protonAda}. The coupled proton spins are much more robust against disturbing fields, due to the homonuclear nature of the coupling. A candidate for the origin of the 
inhomogeneous broadening is the chemical shift anisotropy. It is anisotropic in solids and broadens the NMR spectra of powders \cite{levit05}.

\section{Conclusion}
\label{sec:Conclusion}

We applied a recently developed dynamic mean-field theory for spins (spinDMFT) \cite{graes21} to NMR experiments in order to establish it as a versatile and accurate simulation tool enabling quantitative understanding of microscopic spin physics.

In the first part, we extended spinDMFT to access free-induction decays in homonuclear spin systems. This approach is called non-local spinDMFT (nl-spinDMFT) because it involves calculating non-local quantities that are not directly accessible in spinDMFT. The nl-spinDMFT uses \emph{a priori} computed mean fields to efficiently simulate a quantum cluster of about 10 spins in a mean-field background. The simulation allows us to compute pair correlations, which contribute to the FID. This gives us an understanding of how the macroscopic FID results from microscopic spin dynamics. The extrema in the FID can be traced back order by order to processes involving $n$ spin-spin couplings. 

The agreement of the method with the experimental FID in $\ce{CaF_2}$ \cite{engel74} is excellent for the [100]-direction at short and moderate times. Small deviations at later times are found to be due to less well captured long-range pair correlations, which become relevant at later times. These deviations are larger for the [110]- and [111]-directions because selecting compact clusters with at maximum 9 spins becomes more difficult for higher coordination numbers. In these cases, it may be promising to combine the approach with purely classical or hybrid quantum-classical simulations \cite{elsay15,stark18,stark20}. Our simulations suggest that the choice of the cluster is of great importance. Different pair correlations are accessed by different clusters in nl-spinDMFT. For this reason, we propose to improve the hybrid approach presented in \refcite{stark18,stark20} by computing pair correlations separately considering individual adapted clusters. 

In spin systems with moderate coordination numbers, for example in two-dimensional systems, we expect an excellent performance of nl-spinDMFT. Apart from pair correlations, nl-spinDMFT easily allows one to compute arbitrary quantities on the quantum cluster, which are of interest, for example, for experiments measuring multi-spin correlations \cite{cho05}.

In the second part, we demonstrated the versatility of the mean-field framework by considering the spin dynamics in polycrystalline adamantane \cite{alvar10c}, where motional and powder averaging has to be taken into account. Using spinDMFT, we estimated the proton-spin autocorrelations and used them to calculate the FID and the spin echo of the rare $\ce{^{13}C}$ nuclear spins. While there is some deviation between the numerical and experimental carbon FID, the carbon spin echo is captured very well by the simulation. This suggests the presence of inhomogeneous broadening in the experiment. We account for this in an additional simulation by adding a weak static noise, which leads to an excellent agreement with the experimental FID.

We consider the mean-field framework described by spinDMFT and nl-spinDMFT to be very versatile due to the microscopic nature of the effective mean-field models. For example, spins with $S>\sfrac12$ including quadrupolar interactions can be easily considered in the simulations. Explicit time dependencies can be included in spinDMFT and its extensions without increasing the numerical effort much. This has been demonstrated for the spin echo in adamantane, but more complicated pulse sequences and even pulse imperfections can be considered as well. 

Furthermore, we point out that the nl-spinDMFT can be enhanced by
using CspinDMFT \cite{graes23} in the first step instead of spinDMFT. The second step remains essentially the same. This straightfoward extension is recommended if there is are strong dominating spin interactions which cannot be treated as noise, for instance if dimers of spins with strong intradimer, but weaker interdimer coupling need to be dealt with. Also larger clusters are conceivable as an extension. 

In conclusion, we strongly advocate the use of spinDMFT and its extensions for the prediction and understanding of NMR signals.

\section*{Acknowledgments} 
We acknowledge funding by the Deutsche Forschungsgemeinschaft (DFG) in project UH90/14-1 as well as the DFG for Basic Research in TRR 160. We also thank the TU Dortmund University for providing computing resources on the High Performance Computing cluster LiDO3, partially funded by the DFG in project 271512359. We are grateful for the funding received by T.H. from the DAAD RISE Germany program.


\appendix 

\section{Convergence of pair correlations in calcium fluoride}
\label{app:convergence}

\begin{figure*}
    \centering
    \includegraphics[width=\textwidth]{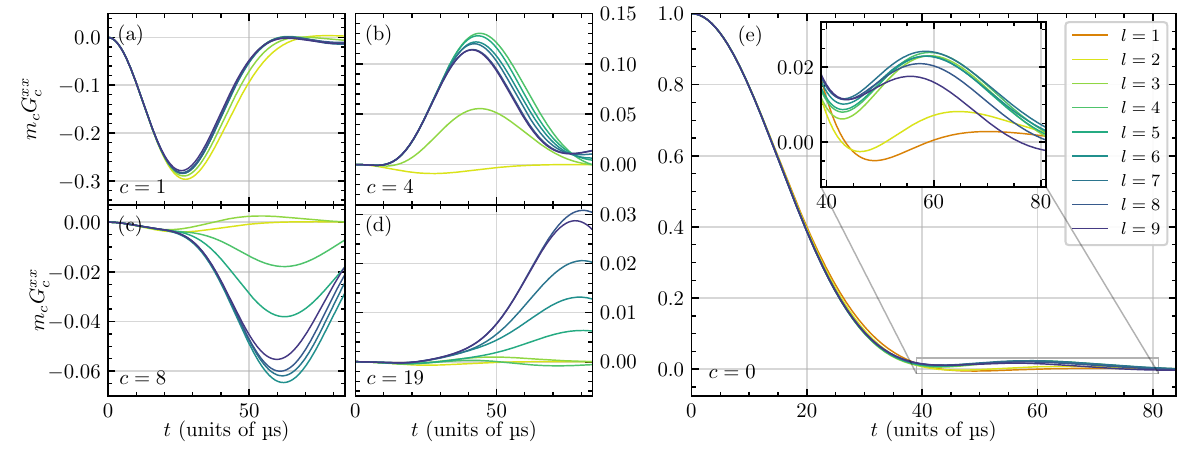}
    \caption{Convergence of the pair correlations from nl-spinDMFT in the [100]-direction with ascending cluster size $l$. Panels (a)-(d) show selected pair correlations with index $c$ according to Tab.~\ref{tab:shelllist} and panel (e) shows the autocorrelation. Pair correlations do not exist for $l~=~1$ which is why the orange line can only be found in panel (e).}
    \label{fig:CaFCONV100}
\end{figure*}

With the help of the algorithm presented in \cref{subsec:choiceofclusters}, we add the most important intermediate spins inducing indirect couplings order by order to the cluster. Comparing the simulation results for different cluster sizes $l$ allows us to track the convergence.
We perform this comparison for the autocorrelation and some selected relevant pair correlations. The result is shown in \cref{fig:CaFCONV100}. As can be seen in panel (e), the autocorrelation changes only very slightly as $l$ is increased. The result for $l=9$ is barely distinguishable from $l=1$ (spinDMFT), except for a tiny maximum at about $\SI{60}{\micro\second}$. This is a perfect \emph{a posteriori} justification for the applicability of spinDMFT. Such a quick convergence is also obtained for the nearest-neighbor pair correlation $c=1$, see panel (a). However, as the pair distance is increased, the convergence becomes slower and slower. 

\begin{figure}
    \centering
    \includegraphics[width=\shapeplotwidth]{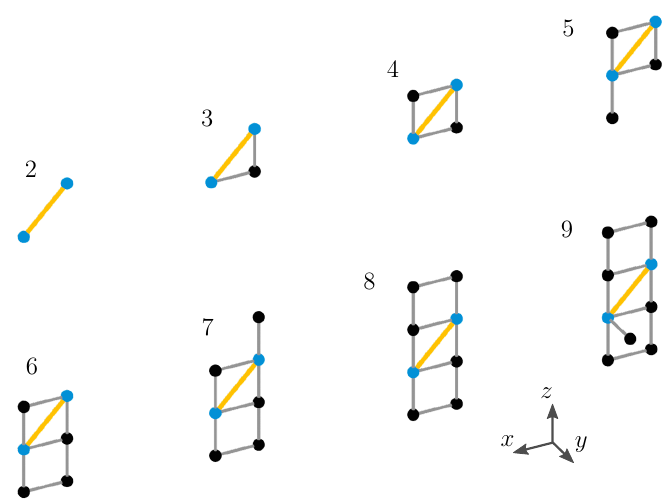}
    \caption{Obtained clusters for the pair correlation with index $c=4$, depicted in orange, for different cluster sizes $l$.}
    \label{fig:shapes_c4}
\end{figure}

Often, we obtain considerable leaps upon adding a single spin, because this spin allows the pair correlation to build up along an additional important path on the lattice. For $c=4$, see panel (b), the third and fourth spin are very important. As can be seen in \cref{fig:shapes_c4}, these are the common neighbors of the two spins directly involved in the pair correlation. The two indirect couplings via spin No.~3 and 4 are the main contributions to this pair correlation. The same behavior can be observed in pair correlation No.~19, see panel (d) and \cref{fig:shapes_c19} for the clusters. Adding the sixth spin enables the first path with four nearest-neighbor links between the two spins of the pair correlation. Each spin that is added until $l=8$ enables another fourth-order path and, thus, increases the fourth-order peak in the pair correlation. Remarkably, the increase is always nearly the same indicating that the effects from different paths roughly adds up linearly in the pair correlation. Studying such effects in detail leads to a much better understanding of the microscopic correlations. Note that with a cluster size of $l=8$ spins, any paths with four nearest-neighbor links are included. Hence, the effect of adding spin No.~9 is significantly smaller. These observations are in perfect accordance with our arguments in \cref{subsec:choiceofclusters,subsec:CaFresults}.

\begin{figure}
    \centering
    \includegraphics[width=\shapeplotwidth]{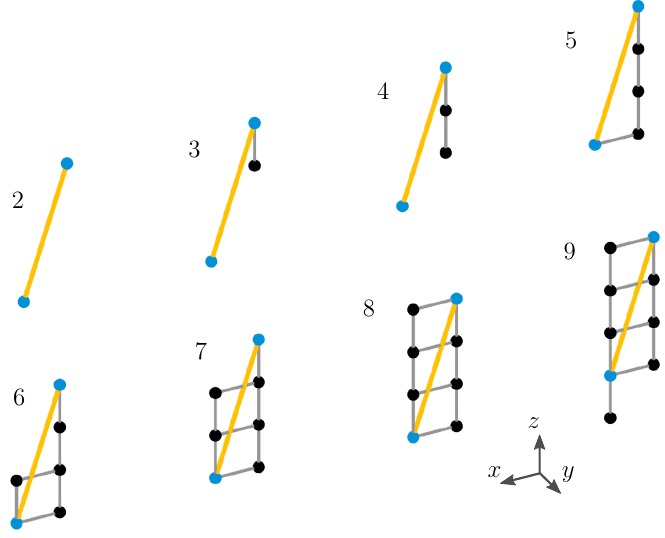}
    \caption{Obtained clusters for the pair correlation with index $c=19$, depicted in orange, for different cluster sizes $l$.}
    \label{fig:shapes_c19}
\end{figure}

The direct understanding of the leaps in the convergence plots allows us to draw conclusions about the accuracy of the computed pair correlations. We consider, for instance, pair correlation No.~19, see \cref{fig:CaFCONV100} (d), to be fairly well captured for $l~\geq~8$, 
because all paths with four nearest-neighbor links are included. Hence, the lowest, most important order is completely accounted for. The clusters for $l=8$ and $l=9$ may be referred to as \emph{compact}. For several other pair correlations with comparably large pair distances this is not the case, see for example $c=11,12,17,18$ in \cref{fig:sh0-19}. Here, the lowest orders are only included partly for $l=9$ which makes the results less reliable. We conclude that the larger the pair distance, the less compact are the clusters at given cluster size and, hence, the less accurate are the resulting pair correlations. Note that the absolute magnitude of the pair correlations decreases as the pair distance increases so that it is justified to truncate at a certain range.

Finally, we also present the convergence of the FID with the cluster size in Fig.~\ref{fig:CaFFID100}. The nl-spinDMFT asymptotically approaches the experimental result and we 
expect an even better agreement for $l>9$.

\begin{figure}
  \centering
  \includegraphics[width=\columnwidth]{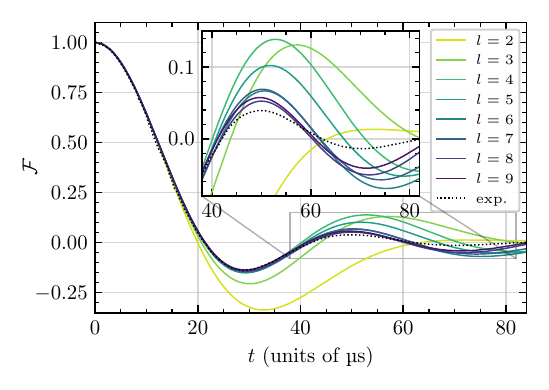}
  \caption{Convergence of the FID from nl-spinDMFT in the [100]-direction with ascending cluster size $l$. The black dotted line corrsponds to the experimental data. }
  \label{fig:shapes_c19}
\end{figure}

\section{FID of $\ce{CaF_2}$ in different directions}
\label{app:otherdir}

\begin{figure*}
    \centering
    \includegraphics[width=\textwidth]{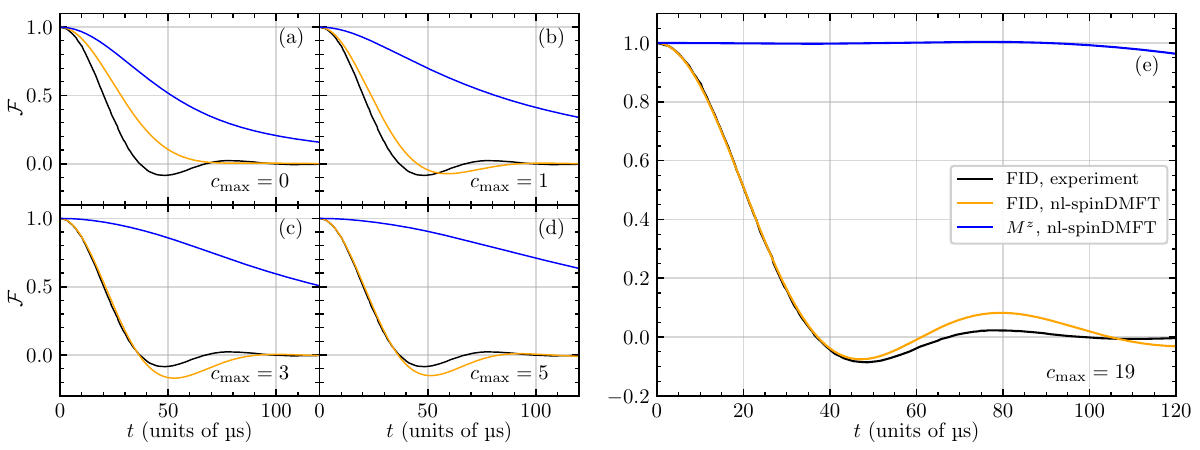}
    \caption{Same as \cref{fig:CaFFID100}, but in the [110]-direction and with $L=8$.}
    \label{fig:CaFFID110}
\end{figure*}

\begin{figure*}
    \centering
    \includegraphics[width=\textwidth]{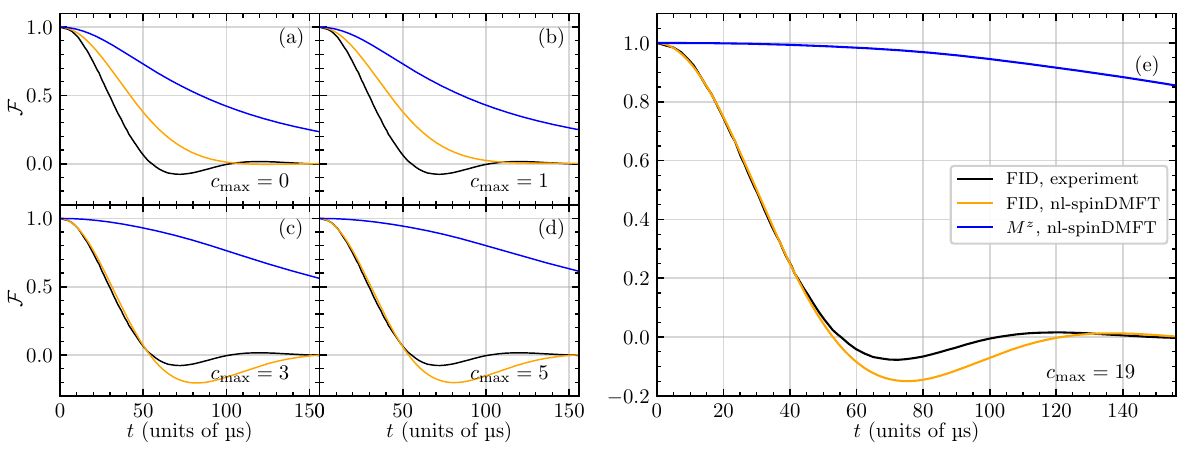}
    \caption{Same as \cref{fig:CaFFID100}, but in the [111]-direction and with $L=14$.}
    \label{fig:CaFFID111}
\end{figure*}

The results for external magnetic field in the [110]- and [111]-directions are shown in \cref{fig:CaFFID110,fig:CaFFID111}. Note that the number of most important couplings $L$ is chosen larger than for the [100]-direction. This is because there are more couplings of similar strength. The time at which nl-spinDMFT becomes inaccurate is about the same as for the [100]-direction (about $\SI{40}{\micro\second}$). In particular, for the [111]-direction the rather late minimum is not captured completely quantitatively. Also, the sum rule of the $z$-direction (blue line) is not well fulfilled for the discussed reasons. The choice of compact clusters requires more and more spins as the coordination number is increased. Since we are limited to 9 spins, the FIDs in [110]- and [111]-direction are less well captured.

\end{document}